\documentclass[12pt]{article}
\makeatletter
\@addtoreset{equation}{section}
\makeatother

\usepackage[pdftex]{graphicx}
\usepackage{subfigure}
\usepackage{array}
\usepackage{amsmath}

\usepackage{rotating}
\usepackage{longtable}
\usepackage{mathtools}
\usepackage{natbib}
\usepackage{makecell}

\usepackage{amsfonts}
\usepackage{array}
\usepackage{geometry}
\usepackage{color}
\usepackage{lipsum}

\usepackage{diagbox} 
\usepackage{pdfpages}
\usepackage{multicol}
\usepackage[utf8]{inputenc}
\usepackage[english]{babel}
\usepackage{mathtools}
\usepackage{bbm}
\usepackage{amssymb}
\usepackage{amsthm}
\usepackage{hyperref}
\usepackage{bm}
\usepackage{mathtools}
\usepackage{fancybox}
\usepackage{multirow}
\usepackage{authblk}
\usepackage{algorithm}
\usepackage{algpseudocode}
\usepackage{setspace}
\usepackage{caption}
\makeatletter
\newsavebox{\@brx}
\newcommand{\llangle}[1][]{\savebox{\@brx}{\(\m@th{#1\langle}\)}%
  \mathopen{\copy\@brx\kern-0.5\wd\@brx\usebox{\@brx}}}
\newcommand{\rrangle}[1][]{\savebox{\@brx}{\(\m@th{#1\rangle}\)}%
  \mathclose{\copy\@brx\kern-0.5\wd\@brx\usebox{\@brx}}}
\makeatother

\newcommand{\vertiii}[1]{{\vert\kern-0.25ex\vert\kern-0.25ex\vert #1\vert\kern-0.25ex\vert\kern-0.25ex\vert}}
\newcommand{\Vertiii}[1]{{\left\vert\kern-0.25ex\left\vert\kern-0.25ex\left\vert #1\right\vert\kern-0.25ex\right\vert\kern-0.25ex\right\vert}}
  
\usepackage{xcolor}
\renewcommand{\raggedright}{\leftskip=0pt \rightskip=0pt plus 0cm}
\title{Variation Pattern Classification of Functional Data}
\author[1]{Shuhao Jiao\thanks{shjiaoqd@gmail.com}}
\author[2]{Ron D.\ Frostig\thanks{rfrostig@uci.edu}}
\author[1]{Hernando Ombao\thanks{hernando.ombao@kaust.edu.sa}}

\affil[1]{Statistics Program,  KAUST, Saudi Arabia}
\affil[2]{Department of Neurobiology and Behavior, UC Irvine, USA}

\date{}
\begin{document}
	\maketitle			
	\setlength\parindent{0pt}
	\setlength{\parskip}{1em}
	\theoremstyle{definition}
	\newtheorem{assumption}{Assumption}
	\newtheorem{theorem}{Theorem}
	\newtheorem{lemma}{Lemma}
	\newtheorem{prop}{ACRosition}
	\newtheorem{definition}{Definition}
	\newtheorem{corollary}{Corollary}
	\newtheorem{remark}{Remark}
\begin{abstract}
A new classification method for functional data is proposed in this paper. This work is motivated by the need to identify features that discriminate between neurological conditions on which local field potentials (LFPs) were recorded. Regardless of the condition, these local field potentials have zero mean and thus the first moments of these random processes do not have discriminating power. We propose the variation pattern classification (VPC) method {which employs the (auto-)covariance operators as the discriminating features} and uses the Hilbert-Schmidt norm to measure the discrepancy between the (auto-)covariance operators of different groups. The proposed VPC method is demonstrated to be sensitive to the discrepancy, {potentially leading to a higher rate of classification}. One important innovation lies in the dimension reduction where the VPC method data-adaptively determines the basis functions (discriminative feature functions) that account for the major discrepancy. In addition, the selected discriminative feature functions provide insights on the discrepancy between different groups because they reveal the features of variation pattern that differentiate groups. Consistency properties are established and, furthermore, simulation studies and the analysis of rat brain LFP trajectories empirically demonstrate the advantages and effectiveness of the proposed method.\\

\noindent{\bf Key words}: Dimension reduction; Discriminative feature function; Functional data analysis; Nearest centroid classifier; (Auto-)covariance operator.
\end{abstract}

\section{Introduction}
\subsection{Motivation from a stroke experiment}
It is important for clinicians to be able to rapidly detect stroke onset in order to minimize the debilitating downstream effects of stroke. Early detection gives the patients the best possible prognosis for quick recovery and minimal neurological damage. In contrast, late detection of stroke is associated with poor prognosis and the patients often take longer time to recover and may not recover from profound effects on motor function, speech and memory. In this paper, we examine local field potentials (LFPs) which are recorded from rats during an experiment. 

The LFPs, regardless of the clinical condition, are random fluctuations around zero. Thus, more interesting and discriminating features  are captured by the second or higher order moments. The motivation of this work comes from the need to identify discriminating features of rat LFPs in a simulated stroke experiment.  
The goals of this paper are the following:
\begin{itemize}
\item[(1)] To develop a statistical method for the discrimination between pre-stroke and post-stroke onset brain signals. In this setting, there is training data available which has known group labels (pre vs.\ post-stroke onset). The goal here is to identify features that best separate the classes of pre-stroke and post-stroke onset signals. 
\item[(2)] To develop a statistical method for classification of signals with unknown group label (normal vs abnormal or stroke in particular). We envision developing a method that can track brain signals online for the purpose of providing some warning for clinicians when the brain signals start to exhibit abnormal features. 
\end{itemize}

\subsection{Our contributions and existing work}
The main contribution of this work is a functional classifier method based on the (auto-)covariance operator under the setting where different groups of functional data have similar mean functions.  The proposed method does not rely on any distributional assumption and thus the classification procedure has broad potential applicability.  It is noted that classification accuracy is influenced by two factors, the discrepancy between groups and background noise in the curve to be classified. 
It is often the case that, as more basis functions are used for discriminating different groups, the discrepancy typically becomes more pronounced. However, variability or uncertainty also increases. Therefore, it is not necessarily advantageous to incorporate many basis functions in discriminant analysis -- if the employed basis functions have low power of discriminating different classes. The proposed VPC method data-adaptively selects the basis functions that account for the major discrepancy between groups. 

In the past two decades, a variety of classification and clustering methods for functional data have been proposed. In \cite{r22}, an extension of linear discriminative analysis to functional data is proposed and relies on a parametric model to reduce the rank. \cite{r31} applied partial least squares in functional linear discriminate analysis. \cite{r23} developed a flexible model-based procedure. \cite{r3} and \cite{r13} applied nearest neighbor rule in functional data classification, and their methods are based on the first moment.  \cite{r28} studied generalized functional linear model, which was used for classification in \cite{r25}. \cite{r26} used multinomial logistic model for multi-class functional data classification. \cite{r5} proposed a functional principal component (FPC) subspace-projected $K$-centers functional discrimination approach. \cite{r6} proposed a correlation-based $K$-centers functional clustering method. \cite{r39} and \cite{r14} employed wavelet methods. \cite{r20} and \cite{r16} proposed discrimination procedures for non-stationary time series. The novelty of these work is that they select the bases from the SLEX library that can best illuminate the difference between two or more classes of processes. \cite{r4} proposed a similar procedure for multivariate non-stationary process. \cite{r33,r34} proposed to select the best discriminative functions from wavelet packets to extract local information for classification problems. \cite{r7} studied a novel functional linear classifier, which is optimal under normality and can be perfect as the number of curves (sample size) diverges. \cite{r8} and \cite{dai} studied the functional Bayesian quadratic classifier. \cite{r36} proposed an interpretable dimension reduction technique for functional data classification. \cite{r21} proposed an algorithm to perform clustering of functional data based on covariance. Some other methods also incorporate the covariance difference, e.g., \cite{r20}, \cite{r5}, and \cite{r14}. However, in these methods, the discrepancy between (auto-)covariance operators is accounted for {by fixed basis functions or }group-wise functional principal components, which do not necessarily capture the discrepancy.

Meanwhile, some existing work for multivariate data also use covariance matrix as a discriminating feature. \cite{r1} and \cite{r17} studied the classification procedure for observations coming from multivariate normal distributions in the case that the two distributions differ both in mean vectors and covariance matrices. \cite{r10} used covariance descriptor in the classification of multivariate data. In brain signal classification, \cite{r9} and \cite{r2} used spatial covariance matrix as a feature. \cite{r35} proposed a frequency-specific spectral ratio statistic and used it as a feature to discriminate different states. \cite{r12} proposed a copula-based algorithm to detect changes in brain signals. These methods use different features to discriminate epochs under different states. However, since the intra-curve (functional) information is not directly incorporated, the methods will not perform well if the discrepancy is mainly present in the variation pattern of trajectories.

Compared to the existing methods, the proposed VPC method has the following advantages: 
1.) The VPC method is entirely data-driven and non-parametric, making it applicable for a broad range of data and robust to potential model misspecification. 2.) The VPC method selects, in a data-adaptive manner, the sequence of orthonormal basis that account for most of the discrepancy of the (auto-)covariance operators. Thus, the selected basis functions improve the classification accuracy and reveal the features with the ability to differentiate groups.
3.) The VPC method takes account of the intra-curve information, {which potentially provides important information discriminating groups}. 4.) The proposed framework can be applied to both independent and dependent functions.

The rest of the paper is organized as follows. Section~\ref{s2} presents the classification procedure, and also displays the theoretical result that the classification is asymptotically perfect under some regularity conditions. In Section~\ref{s4}, we study the finite sample properties of the classifier by simulations. In Section~\ref{s5}, the VPC method is implemented to classify the LFP epochs. Conclusion is made in Section~\ref{s6}. Technical proofs and additional  real data analysis on phoneme data can be found in the supplementary material.

\section{Model, consistency, and algorithm}
\label{s2}
\subsection{General setting and preliminaries} 
Let $\{X_k(t)\colon k\in\mathbb{N},\ t\in [0,1]\}$ be a set of random functions such that the realizations are elements of the Hilbert space $L^2[0,1]$, where the inner product is defined as $\langle x,y \rangle=\int_0^1x(t)y(t)dt$, 
and the norm is defined as $\|x\|^2=\langle x,x\rangle=\int_0^1x(t)^2dt$. Assume $E\|X\|^2<\infty$, and define the mean function by $\mu(t)=E\{X(t)\},$
and the covariance operator $C(\cdot)\colon L^2[0,1]\to L^2[0,1]$ by $C(\cdot)=\mathbb{E}\{\langle X,\cdot\rangle X\}.$
 By the Mercer's theorem, assume $C(\cdot)=\sum\limits_{j=1}^{\infty}\lambda_j\langle v_j,\cdot\rangle v_j,$ where $\{\lambda_j\colon j\in\mathbb{N}_+\}$ are the positive eigenvalues (in strictly descending order) and $\{v_j \colon j \in\mathbb{N}_+\}$ are the corresponding orthonormal eigenfunctions, so that $C(v_j) = \lambda_jv_j$ and $\|v_j\| = 1$.
The Hilbert--Schmidt norm of an operator $\Phi$ is defined as: $\|\Phi\|^2_\mathcal{S}=\sum\limits_{i,j}|\Phi_{i,j}|^{2},$ where $\Phi_{i,j}=\langle \Phi(e_{i}), e_{j}\rangle$, and $\{e_i\colon i\in\mathbb{N}_+\}$ is a sequence of orthonormal basis functions. This norm does not depend on the choice of $\{e_i\colon i\in\mathbb{N}_+\}$.

Consider a sequence of functions in $L^2[0,1]$ for each group $\Pi_g~(g=0,1)$, $X_1^{(g)}(t),\ldots,X_{n_g}^{(g)}(t)$, $g=0,1$ and $n_0+n_1=n$, where $g$ is the group index and {the functions in either group are weakly stationary}. Define the mean function and the (auto-)covariance operator at lag $h$ as $E\{X_k^{(g)}(t)\}=\mu(t)$ and
$C^{(h)}_g(\cdot)=E\{{X_k^{(g)}\langle X_{k+h}^{(g)},\cdot\rangle}\},~C^{(-h)}_g(\cdot)=E\{X_{k+h}^{(g)}\langle X_{k}^{(g)},\cdot\rangle\},$
where $h\in\mathbb{N}$. In practice, $\{C^{(h)}_g(\cdot)\colon h\in\mathbb{N},~g=0,1\}$ are unknown, and are estimated by the following estimators 
$\widehat{C}^{(h)}_g(\cdot)=(n_g-h)^{-1}\sum\limits_{k=1}^{n_g-h} \langle X^{(g)}_{k+h},\cdot\rangle X^{(g)}_k,~\widehat{C}^{(-h)}_g(\cdot)=(n_g-h)^{-1}\sum\limits_{k=1}^{n_g-h} \langle X^{(g)}_{k},\cdot\rangle X^{(g)}_{k+h}.$

It is assumed that the mean functions of the two groups are both zero, say, $\mu(t)=0$. This is a reasonable assumption for many brain signals (e.g., electroencephalograms, local field potentials, magnetoencephalograms), where recordings always fluctuate around the zero line. Here we consider the cases where groups share the same mean. If the means are the same but not zero, discriminating (lagged) second moment operators is equivalent to discriminating (auto-)covariance operators due to the fact $C_g^{(h)}(\cdot)=E\{X_{k}^{(g)}\langle X_{k+h}^{(g)},\cdot\rangle\}-\mu\langle\mu,\cdot\rangle$. Thus the assumption $\mu(t)=0$ does not lead to any loss of generality. When the mean functions of different groups are not identical but similar, the VPC classifier can still give decent classification results (see the analysis of phoneme data in the supplementary material).

\subsection{Discriminative feature functions}
\label{s3.1}
A widely-accepted metric to evaluate the difference between the two covariance operators $C_0^{(0)}$ and $C_1^{(0)}$ is the Hilbert-Schmidt metric $\|\cdot\|_\mathcal{S}$, defined as 
$$\|C_0^{(0)}-C_1^{(0)}\|{^2}_\mathcal{S}=\sum_{i,j=1}^{\infty}\langle(C_0^{(0)}-C_1^{(0)})(\nu_i),\nu_j\rangle^2.$$
Note that the evaluation of the metric does not rely on the selection of the orthonormal basis $\{\nu_j\colon j\ge1\}$. However, it is important to select the basis that captures most of the discrepancy due to the decision rule which is now briefly described. Assume that $\{X_k^{(g)}(t)\colon k\in\mathbb{N}\}$ are the observed functions for group $g$. Suppose $Y(t)\in L^2[0,1]$ is a new function whose group membership is to be determined. Notationally, let $\mathcal{Y}(\cdot)(t)=\langle Y,\cdot\rangle Y(t)$ be a one-dimensional compact operator driven by $Y(t)$. Our lag-$0$ centroid classifier assigns $Y$ to group $\Pi_g$ if 
$$D_d(\mathcal{Y},C^{(0)}_g)<D_d(\mathcal{Y},C^{(0)}_{1-g}),$$ where $D_d$ is the Hilbert-Schmidt metric distance determined by $d$ orthonormal basis functions $\{\nu_j(t)\colon j=1\ldots,d\}$, say, 
$D_d(\mathcal{Y},C^{(0)}_g)=\sum\limits_{i,j=1}^{d}\langle (Y\langle Y,\nu_i\rangle-C^{(0)}_g(\nu_i)),\nu_j\rangle^2.$

Clearly $C^{(0)}_{0}-C^{(0)}_{1}$ is compact, and suppose the $C^{(0)}_{0}-C^{(0)}_{1}$ admits the compact spectral decomposition $(C^{(0)}_{0}-C^{(0)}_{1})(\cdot)=\sum\limits_{i=1}^{\infty}\lambda_{0,i}\langle \nu_{0,i},\cdot\rangle \nu_{0,i}$, and accordingly $(C^{(0)}_{0}-C^{(0)}_{1})^2(\cdot)=\sum\limits_{i=1}^{\infty}\lambda_{0,i}^2\langle \nu_{0,i},\cdot\rangle \nu_{0,i}$.
By the nature of Hilbert-Schmidt norm, $\|C^{(0)}_{0}-C^{(0)}_{1}\|^2_\mathcal{S}=\sum\limits_{i=1}^\infty\lambda_{0,i}^2.$
Therefore, most discrepancy is captured by the eigenfunctions associated with the large eigenvalues of the positive-definite operator $(C^{(0)}_{0}-C^{(0)}_{1})^2(\cdot)$, and we propose to use the eigenfunctions associated with the large eigenvalues of the compact symmetric operator $C(\cdot)=(C^{(0)}_{0}-C^{(0)}_{1})^2(\cdot)$, which are called as the discriminative feature functions. {In the cases where the discrepancy cannot be explained by group-wise functional principal components $\{\phi_d^{(g)}\colon d\ge1,g=0,1\}$, say, $\langle\nu_{0,i},\phi^{(g)}_{d}\rangle=0$ for all $i,~d$ and $g=0,1$, the classifier based on group-wise functional principal components cannot yield satisfactory results. In comparison, the discriminative feature functions are extracted from the difference operator $(C_0^{(0)}-C_1^{(0)})^2$ and can always capture the discrepancy between the covariance operators.} 

Similarly, as for the auto-covariance operators, the eigenfunctions of $(C_0^{(h)}+C_0^{(-h)}-(C_1^{(h)}+C_1^{(-h)}))^2$ associated with its large eigenvalues account for most of the discrepancy $C_0^{(h)}+C_0^{(-h)}-(C_1^{(h)}+C_1^{(-h)})$. To find the most important discriminative basis functions, select the first $d_h$ eigenfunctions of the positive definite operator 
$\mathcal{R}_h=(C_0^{(h)}+C_0^{(-h)}-(C_1^{(h)}+C_1^{(-h)}))^2,$
associated with the first $d_h$ largest eigenvalues of $\mathcal{R}_h$. The estimator of $\mathcal{R}_h$ is $\widehat{\mathcal{R}}_h=(\widehat{C}_0^{(h)}+\widehat{C}_0^{(-h)}-(\widehat{C}_1^{(h)}+\widehat{C}_1^{(-h)}))^2.$ In the following of the paper, $\{\nu_{h,j}(t)\colon j=1,\ldots,d_h\}$ denote the selected discriminative feature functions for the lag-$h$ (auto-)covariance operator.

\textit{Remark}.
Consider LFP traces and represent the trajectories with the following Fourier expansion,
$X_k^{(g)}(t)=\sum\limits_{j=1}^{L} \left(a^{(g)}_{kj}\cos(jt)+b^{(g)}_{kj}\sin(jt)\right),$
where $L$ is the length of each epoch. As the number of epochs increases, more basis functions can be incorporated to discriminate $\{a_j^{(0)},b_j^{(0)}\}$ and $\{a_j^{(1)},b_j^{(1)}\}$ over a wider range of frequencies $j$. If the data of different groups present discrepancy over a wide range of frequency bands, the classification can approach perfectness as the sample size increases. {Note that this expansion was also used in the context of topological data analysis (see e.g.~\cite{r40}).}

\subsection{Classification procedure}
\label{s3.4} 
Assume that $\{Y_k(t)\in L^2[0,1]\colon k=1,\ldots, p{+1}\}$ are consecutively collected from the same group, whose group memberships are to be predicted or determined jointly. {$p$ signifies the maximal incorporated lag}. In some cases it might be unrealistic to assume that the consecutive samples $\{Y_k(t)\colon k=1,\ldots, p{+1}\}$ come from the same class. However, in our application, the LFP signals remained abnormal after the stroke-onset for a sufficient long time, so that {it is possible} to collect consecutive functions in the post-stroke onset state. Let ${\kappa}^{(h)}_g=C_g^{(h)}+C_g^{(-h)}$, ${\hat{\kappa}^{(h)}_g=\widehat{C}_g^{(h)}+\widehat{C}_g^{(-h)}}$ and
\begin{align*}
\widehat{\kappa}_{y,h}(\cdot)=\frac{1}{p+1-h}\sum_{k=1}^{p+1-h}Y_{k}\langle Y_{k+h},\cdot\rangle+\frac{1}{p+1-h}\sum_{k=1}^{p+1-h}Y_{k+h}\langle Y_{k}.\cdot\rangle,
\end{align*}
In addition, define $\{\hat{\nu}_{h,j}\colon j\ge1\}$ to be the empirical discriminative feature functions obtained from 
$\widehat{\mathcal{R}}_h=(\widehat{C}_0^{(h)}+\widehat{C}_0^{(-h)}-(\widehat{C}_1^{(h)}+\widehat{C}_1^{(-h)}))^2.$
The (auto-)covariance operators at different lags may have varying levels of discriminating power, so we consider the weighted classifier. The procedure is summarized in Algorithm~\ref{alg3}.
\begin{algorithm}[htb]
\caption{Classification algorithm}
\label{alg3}
\textbf{ Step 1}. Fix $d_h$, obtain the eigenfunction of $\widehat{\mathcal{R}}_h$, say, $(\widehat{\nu}_{h,j}\colon j=1,\ldots,d_h)$. 

\textbf{ Step 2}. Compute the scores $\widehat{S}_{g,ij}^{h}=\langle\widehat{\kappa}^{(h)}_{g}(\widehat{\nu}_{h,i}),\widehat{\nu}_{h,j}\rangle$ for $i,j=1,\ldots,d_h$.

\textbf{ Step 3}. Compute $$\widehat{D}_g=\sum_{h=0}^pW(h)\sum^{d_h}_{i,j=1}(\widehat{S}_{g,ij}^h-\langle\widehat{\kappa}_{y,h}(\widehat{\nu}_{h,i}), \widehat{\nu}_{h,j}\rangle)^2.$$ If $\widehat{D}_0-\widehat{D}_1<0$, classify $Y_1,\ldots,Y_{p+1}$ to $\Pi_0$, otherwise, classify them to $\Pi_1$. 
\end{algorithm}

\textit{Remark}.
{Besides (auto-)covariance operator, other operators which capture the second moment information can also be incorporated (e.g., long-run covariance operator).}

The value $\sum^{d_h}\limits_{i,j=1}(\widehat{S}_{g,ij}^h-\langle\widehat{\kappa}_{y,h}(\widehat{\nu}_{h,i}), \widehat{\nu}_{h,j}\rangle)^2$ can be small just because the amplitude of the lag-$h$ (auto-)covariance is small, so they are scaled by $W(h)$. The weight $W(h)$ depends on two factors: 1).\ The amplitude of the (auto-)covariance operator $\|\widehat{C}_0^{(h)}\|_\mathcal{S}+\|\widehat{C}_1^{(h)}\|_\mathcal{S}$, and 2).\ The classification rate based on lag $h$ only, $P(h)$. {The explicit expression of $P(h)$ cannot be obtained here since no assumptions on distribution are made. However, an empirical version of $P(h)$ can always be obtained by classifying the training set based on lag $h$ only through some cross-validation procedure.} In our application, $W(h)=(\|\widehat{C}_0^{(h)}\|_\mathcal{S}+\|\widehat{C}_1^{(h)}\|_\mathcal{S})^{-1}\exp(\alpha P(h))$, where $\alpha$ is a tuning parameter and can be fixed by a cross-validation (CV) procedure. {Here we briefly describe a feasible procedure (Monte-carlo CV) to select $\alpha$, the maximal lag $p$ and $P(h)$. A sequence of consecutive functions are randomly selected from the samples as the testing set, and use the rest samples to train the classifier based on single lag $h$ and obtain the empirical classification rate of two groups. For a sufficient range of $h$, repeat the procedure multiple times to obtain the average rate (over repetitions and groups) and employ it as $P(h)$. Then apply the same procedure to obtain the empirical classification rates of the whole trained classifier that incorporates all $p$ lags, where a set of pre-specified candidate tuning parameters $p,\alpha$ are tested. The resampling and classification steps are repeated multiple times to calculate the average classification rates under different pairs of $p,\alpha$, and the tuning parameters corresponding to the largest average classification rate shall be selected.}

\textit{Remark}.
The log-Euclidean metric and the affine invariant Riemannian metric are commonly used for covariance matrices. However, the matrix logarithm cannot be extended to infinite-dimensional trace-class functional operators. The eigenvalues of trace-class operator typically decay to zero, making it difficult to extend these distance to functional data. In contrast, the distance induced by Hilbert-Schmidt norm is well defined for (auto-)covariance operator and can also produce reasonable between-group comparison. This point is also discussed in \cite{r30}.

\textit{Remark}.
As a side remark, when classifying the auto-covariance operators, one major challenge is the unbalanced variance of $X(t)X(s)$ across groups. Pronounced unequal level of variance leads to unbalanced classification, necessitating additional steps (e.g., scaling the functions by their norm, and pre-classify by the amplitude). Scaling can make the variability of different groups at the same level and help yield balanced classification, but may lead to the loss of classification power if the amplitude is informative of discrimination. If the variability of one group is significantly higher than that of the other one, we can do pre-classification based on the amplitude. {Specifically, first set a threshold $\tau$, which can be selected by cross-validation, given a curve $Y(t)$ to be classified, if $\|Y\|>\tau$, classify it into the group with higher variation level, and the rest functions are scaled to unit norm and classified by the proposed method.} 

\subsection{Classification among multiple groups}
Assume we have $G$ groups $\Pi_1,\ldots,\Pi_G$ of functions, where $G>2$, the VPC procedure can be easily extended to this case. In the multi-class classification, we propose to do pairwise classification for different pairs of groups. More specifically, we first discriminate the first two groups $\Pi_1$ and $\Pi_2$, if the curves with unknown group label are classified into $\Pi_1$, then we further do pairwise comparison between $\Pi_1$ and $\Pi_3$. The procedure is finished when all necessary pairwise discrimination are conducted. Formally, the predicted index is defined as $\hat{g}=\arg\min_{g=1,\ldots,G}\widehat{D}_g.$

\subsection{Consistency}
Now we discuss the consistency property of the classifier. By the Mercer's theorem, assume that
$\mathcal{R}_h(\cdot)=\sum\limits_{j=1}^{\infty}\lambda_{hj}\langle \nu_{h,j},\cdot\rangle \nu_{h,j},$
where $\lambda_{h1}>\lambda_{h2}>\ldots$. Notationally, let $\sigma_{ij}^h=E\{y_{ij}^h-\langle\kappa_g^{(h)}(\nu_{h,i}),\nu_{h,j}\rangle\}^2$ and $y_{ij}^h=\langle\hat{\kappa}_{y,h}(\nu_{h,i}),\nu_{h,j}\rangle$. The following theorem gives an upper bound of the misclassification rate of the classifier with the known discriminative feature functions.

\begin{theorem}{Theorem 1.}{}
Assuming there exists at least one $h$ such that $\|\mathcal{R}_{h}\|_\mathcal{S}>0$ and $E\|X^{(g)}_k\|_4<\infty$ for $g=0,1$, the misclassification rate satisfies
$$P(\Pi_{1-g}\vert\ \Pi_{g}, \{\nu_{h,j},\colon h\ge0,j\ge1\})\le \frac{4\sum\limits_{h=0}^p\left(W(h)\sum\limits_{i,j=1}^{d_h}\sigma_{ij}^h\right)}{\sum\limits_{h=0}^p\left(W(h)\sum\limits_{j=1}^{d_h}\lambda_{hj}\right)}\wedge 1.$$
\end{theorem}

Theorem~1 presents an upper bound of the misclassification rate. The numerator is the variance, and the denominator is the discrepancy evaluated by the discriminative feature functions. The corollary below follows from Theorem~1 consequently.
\begin{corollary}{Corollary 1.}{}%
\label{co1}
Assume that the conditions in Theorem~1 hold, and if
 $${\sum\limits_{h=0}^{p}\left(W(h)\sum\limits_{j=1}^{d_h}\lambda_{hj}\right)}\bigg/{\sum\limits_{h=0}^p\left(W(h)\sum\limits_{i,j=1}^{d_h}\sigma_{ij}^h\right)}\to\infty,$$
the classification tends to be perfect as $p\to\infty$.
\end{corollary}
This corollary states that if the (auto-)covariance operators at all lags are sufficiently discriminative, specifically, the discrepancy is large enough compared to the variance, the misclassification rate goes to zero if the number of incorporated lags increases to infinity.

The problem is that the discriminative feature functions $\{\nu_{h,j},\colon h\ge0,j\ge1\}$ and $\{\kappa_g^{(h)}\colon h\ge0, g=0,1\}$ are unknown and should be estimated from the samples, and it is impossible to incorporate all lags. If a huge amount of discriminative basis functions and lags were incorporated, the estimation error could reduce the classification power. 
However, under some regularity conditions, the perfect classification can still be achieved as the sample size goes to infinity. To establish the consistency of the estimated classifier, several assumptions are needed and introduced below.
\begin{itemize}
\item[({\bf {\bf A1}})]For each $0\le h\le p_n$, $\mathcal{R}_h(\cdot)=\sum\limits_{j\ge1}\lambda_{hj}\langle\nu_{h,j},\cdot\rangle\nu_{h,j}$, where $\lambda_{h1}>\lambda_{h2}>\cdots$ and $\sum_{j}\lambda_{hj}<\infty$, and there exists $\alpha_h>2$, so that $\delta_{hj}=\lambda_{hj}-\lambda_{h,j+1}\ge \mbox{const.}j^{-\alpha_h}$. ``$\mbox{const.}$'' signifies a constant not related to $h$.
\item[({\bf {\bf A2}})] $p_n,d_{h,n}$ are selected so that $p_n/n\to0$, $d_{h,n}=n^{1/\tau_h}$, and $$\sum\limits_{h=0}^{p_n}W(h)n^{(2\alpha_h+1)/\tau_h-1}\to0,\qquad \sum\limits_{h=0}^{p_n}W(h)n^{1/\tau_h-1/2}\to0.$$
\item[({\bf {\bf A3}})] The process $\{X^{(g)}_k(t)\colon k\ge1\}$ is weakly dependent so that $E\|\widehat{C}^{(h)}_g-C^{(h)}_g\|^{2q}\le O((n_g-h)^{-q})$ for any $h$, $g$ and $q=1,2$. $n_0/n_1\to\eta$ where $\eta\ne0,\infty$.
\end{itemize}

Assumption ({\bf A1}) guarantees the identifiability of the discriminative feature functions of each lag $h$, and $\alpha_h>2$ comes from $\sum_{j}\lambda_{hj}<\infty$. Assumption ({\bf A2}) assures the selection of $p_n$ and $d_{h,n}$ not leading to overly large estimation error. In assumption ({\bf A3}), \cite{r18} have shown $E\|\widehat{C}^{(0)}_g-C^{(0)}_g\|^{2}=O(n^{-1})$ for $L^4-m$-approximable process, and similar arguments can be employed to extend this result to the cases $h\ne0$, and $E\|\widehat{C}^{(h)}_g-C^{(h)}_g\|^{2q}\le O((n_g-h)^{-q})$ always holds for $m$-dependent processes as $q=1,2$.

\begin{theorem}{Theorem 2.}{}%
Assuming that ({\bf A1})---({\bf A3}) and the assumptions in Theorem~1 hold, if 
$${\sum\limits_{h=0}^{p_n}\left(W(h)\sum\limits_{j=1}^{d_{h,n}}\lambda_{hj}\right)}\bigg/{\sum\limits_{h=0}^{p_n}\left(W(h)\sum\limits_{i,j=1}^{d_{h,n}}\sigma_{ij}^h\right)}\to\infty,$$
then $P(\Pi_{1-g}\vert\ \Pi_g, \{\hat{\nu}_{h,j},\colon h\ge0,j\ge1\})\overset{p}\to0$, as $n\to\infty$.
\end{theorem}

Theorem~{2} illustrates that under some regularity conditions, the classification based on the estimated discriminative feature functions still approaches perfectness if the discrepancy between two groups is sufficiently large.

\section{Simulation}
\label{s4}
\subsection{Classification of functional moving average processes}
To study the finite sample performance of the proposed method, we employed the method to classify two FMA(3) processes, which follow the recursive equation $X^{(g)}_{k}(t)=\epsilon_{k}(t)+ \sum\limits_{\ell=1}^3\psi_{g\ell}(\epsilon_{k-\ell})(t),$
where $\{\epsilon_k(t)\colon k\ge1\}$ are zero-mean $i.i.d$ innovation functions across $k$. All the functions were simulated with the first 21 Fourier basis functions, here denoted as a functional vector 
$\bm{F}(t)=(F_1(t),\cdots,F_{21}(t))'.$
Two templates $\Psi_g(\cdot)$, $g=0,1$ were created for the coefficient operators, specifically, $\psi_{g\ell}(\cdot)=a_{\ell}\Psi_g(\cdot)$. Here we set $a_\ell=0.4$.
The template coefficient operators admits the basis representation $\Psi_g(\cdot)=\bm{F}'K_g\langle\bm{F},\cdot\rangle$, where $K_g$ is a $21\times 21$ matrix and each element was sampled from the normal distribution $K_{g,ij}\sim\mathcal{N}(0,(\sigma'_g\sigma_g)_{ij})$, where
\begin{align*}
\sigma_0=(1, \bm{1}_5\otimes(0.8,0.8,1,1)),\qquad\sigma_1=(1,\bm{1}_5\otimes(1,1,0.8,0.8)),
\end{align*}
and $\bm{1}_5=(1,1,1,1,1)$. 50, 100, or 600 functions were simulated for each group as the training set, and another 100 functions were simulated for the testing set to obtain the empirical $P(h)$ and select $\alpha$ {by the CV procedure described in Section 2.3}. The dimension $d_h$ is selected as the smallest dimension satisfying $\sum\limits_{j=1}^{d_h}\lambda_{hj}/\sum\limits_{j\ge1}\lambda_{hj}\ge 90\%$.

The maximal lag $p$ under consideration were 0,1,2,3,4. 
Under each setting, 100 sets of new functions ($p+1$ consecutive simulated functions in each set) were classified and the classification rates were computed for comparison. Under each simulation setup, the simulation runs were repeated for 200 times and the average classification rates and the standard error of the empirical classification rates are displayed in Table~\ref{FMA}. Here, we set $W(h)=(\|\widehat{C}_0^{(h)}\|_\mathcal{S}+\|\widehat{C}_1^{(h)}\|_\mathcal{S})^{-1}\exp(10 P(h))$.

The optimal classification performance was typically reached at $p= 3$. That is because the discrepancy between the two FMA processes concentrates on the first 3 lags, and two curves become independent as the lag is greater than 3. 
\begin{table}[!h]
\small
	\centering
	\caption{Average classification rates and the corresponding standard errors (in the parenthesis)}
	\begin{tabular}{p{0.3in}p{0.3in}p{0.4in}p{0.4in}p{0.4in}p{0.4in}p{0.4in}p{0.4in}}
	\hline\hline
\multicolumn{2}{c}{\diagbox{$p$}{$n$}} & \multicolumn{2}{c}{50} & \multicolumn{2}{c}{100}& \multicolumn{2}{c}{600}\\
	\hline
        \multicolumn{2}{c}{\multirow{3}{*}{0}}   & $\Pi_0$ & $\Pi_1$ & $\Pi_0$ & $\Pi_1$ & $\Pi_0$ & $\Pi_1$\\
        \cline{3-8}
                      &        &   0.826   & 0.807 &0.889 &0.872 & 0.950&0.946 \\
                      &         &    (0.071) & (0.068) &(0.051) &(0.058) & (0.028)&(0.032)\\
	\hline
        \multicolumn{2}{c}{\multirow{3}{*}{1}}   & $\Pi_0$ & $\Pi_1$ & $\Pi_0$ & $\Pi_1$ & $\Pi_0$ & $\Pi_1$\\
         \cline{3-8}
                       &           & 0.843 & 0.827 & 0.907&0.892 & 0.962& 0.962\\
                       &              & (0.062) &(0.057) &(0.042) &(0.047) & (0.022)&(0.023) \\
	\hline
        \multicolumn{2}{c}{\multirow{3}{*}{2}}  & $\Pi_0$ & $\Pi_1$ & $\Pi_0$ & $\Pi_1$ & $\Pi_0$ & $\Pi_1$\\
          \cline{3-8}
       		     &              & 0.847 &0.834 & 0.909&0.899 & 0.964& 0.964\\
		     &              & (0.062) & (0.055)& (0.041)&(0.045) & (0.020)&(0.023) \\
	\hline
	\multicolumn{2}{c}{\multirow{3}{*}{3}}  & $\Pi_0$ & $\Pi_1$ & $\Pi_0$ & $\Pi_1$ & $\Pi_0$ & $\Pi_1$\\
          \cline{3-8}
                      &              & 0.850 & 0.839&0.912 &0.904 & 0.966& 0.966\\
                      &              & (0.062) &(0.057) &(0.040) &(0.043) & (0.018)& (0.021)\\
	\hline
	\multicolumn{2}{c}{\multirow{3}{*}{4}}  & $\Pi_0$ & $\Pi_1$ & $\Pi_0$ & $\Pi_1$ & $\Pi_0$ & $\Pi_1$\\
          \cline{3-8}
                      &              & 0.848 &0.838 & 0.912&0.903 & 0.966& 0.966\\
                      &              & (0.062) &(0.056) &(0.040) &(0.043) & (0.018)& (0.022)\\
	\hline
	\hline
	\end{tabular}
	\label{FMA}
\end{table}

\textit{Remark}. {Although the functions were simulated under Gaussianity in the simulation, the VPC classifier does not rely on any distributional assumptions and thus also works for other distributions.}

\subsection{Comparison with other methods}
{It is hard to conduct fair comparison with the methods based on other features (e.g., spectrum and dependence structure of univariate time series), because if the employed feature is more informative of the discrepancy, the corresponding method should be superior to others. Thus, in the comparison, only functional methods which incorporate second moment information are considered.}  Three methods were considered: 1) the proposed method (VPC); 2) projection method (denoted by PJ, \cite{r5}); 
3) functional quadratic classifier (denoted by FQC, \cite{r8}, \cite{dai}).  Since the competitor methods do not incorporate the dependence information across curves, we only consider lag 0 in the comparison.
Therefore we simulated two classes ($n_1=n_2=200$) of independent functions with $24$ B-spline basis functions $\{B_1(t),\ldots,B_{24}(t)\}$ over the unit interval $[0,1]$. 
The functions in each group admit the following basis representation
$X_k^{(g)}(t)=\sum_{j=1}^{24}\xi^{(g)}_{kj}B_j(t)$, $g=0,1$.

The scores of two groups follow two different normal distributions respectively $\{\xi^{(g)}_{kj}\colon{j=1,\ldots,24}\}\sim\mathcal{N}(0,\sigma^b_g)$, where 
\begin{align*}
\sqrt{\sigma^b_0}=I_3\otimes\mbox{diag}(b,b,a,b,b,b,b,b),\sqrt{\sigma^b_1}=I_3\otimes\mbox{diag}(b,b,b,b,b,a,b,b),
\end{align*}
$I_3$ is the $3\times 3$ identity matrix and $a$, $b$ are two constants satisfying $a^2+7b^2=100$. Clearly, a larger value of $a$ leads to more substantial discrepancy between the two groups. 
We simulated 200 curves for each group and then split these curves into 100 curves for training and 100 for testing. The classification procedure was conducted 
200 times. The average classification rates (ACR) for each group are presented in Table~\ref{compare}, where the standard errors of the empirical classification rates are shown in the parenthesis. 
The simulation results demonstrate that the projection method and the functional quadratic classifier are inferior to the VPC method and sometimes struggle to distinguish between the groups even though 
they also incorporate the covariance operator in the classification procedure, especially when the discrepancy is not substantial.
One explanation for this suboptimal performance is that the methods use group-wise principal components which are not guaranteed to capture the discrepancy, that is, the major functional principal components in the different groups might be very similar and do not capture the discrepancy between the two groups. In contrast, the VPC method utilizes the feature functions that account for most of the {discrepancy} between the covariance operators, leading to the ability to detect the discrepancy more {\color{black}effectively and efficiently. 
\begin{table}[!h]
\small
	\centering
	\caption{Average classification rates and the corresponding standard errors (in the parenthesis) of the three methods}
	\begin{tabular}{ccccccc}
	\hline
	\hline
	\multicolumn{7}{c}{$a^2=20$}\\
	\hline
\multicolumn{1}{c}{Methods} & \multicolumn{2}{c}{VPC} & \multicolumn{2}{c}{PJ}& \multicolumn{2}{c}{FQC}\\
	\hline
         Group  & $\Pi_0$ & $\Pi_1$ & $\Pi_0$ & $\Pi_1$ & $\Pi_0$ & $\Pi_1$\\
         \hline 
         ACR & 0.55(0.08) &0.55(0.08) & 0.53(0.11) & 0.52(0.11) & 0.95(0.03)& 0.05(0.03) \\
	\hline
	\multicolumn{7}{c}{$a^2=40$}\\
	\hline
\multicolumn{1}{c}{Methods} & \multicolumn{2}{c}{VPC} & \multicolumn{2}{c}{PJ}& \multicolumn{2}{c}{FQC}\\
	\hline
         Group  & $\Pi_0$ & $\Pi_1$ & $\Pi_0$ & $\Pi_1$ & $\Pi_0$ & $\Pi_1$\\
         \hline 
         ACR & 0.78(0.09)& 0.79(0.09) & 0.68(0.10)& 0.70(0.10) & 0.94(0.03)& 0.25(0.11) \\ 
	 \hline 
	\multicolumn{7}{c}{$a^2=60$}\\
	\hline
\multicolumn{1}{c}{Methods} & \multicolumn{2}{c}{VPC} & \multicolumn{2}{c}{PJ}&\multicolumn{2}{c}{FQC}\\
	\hline
         Group  & $\Pi_0$ & $\Pi_1$ & $\Pi_0$ & $\Pi_1$ & $\Pi_0$ & $\Pi_1$\\
         \hline 
         ACR & 0.89(0.08)& 0.88(0.08) & 0.85(0.06)& 0.84(0.006) &  0.95(0.03)& 0.55(0.15) \\
	\hline
	\multicolumn{7}{c}{$a^2=80$}\\
	\hline
\multicolumn{1}{c}{Methods} & \multicolumn{2}{c}{VPC} & \multicolumn{2}{c}{PJ}&\multicolumn{2}{c}{FQC}\\
	\hline
         Group  & $\Pi_0$ & $\Pi_1$ & $\Pi_0$ & $\Pi_1$ & $\Pi_0$ & $\Pi_1$\\
         \hline 
         ACR & 0.94(0.07) & 0.95(0.07) & 0.96(0.03)& 0.96(0.03) & 0.98(0.03)& 0.81(0.18) \\
	 \hline 
	 \hline
	\end{tabular}
	\label{compare}
\end{table}
\subsection{Discriminative feature functions}
To show the necessity of selecting the basis function with the most discriminative power in a more straightforward manner, and how the discriminative feature functions discriminate different groups, we simulated two groups of independent functions with both common and uncommon components. 200 functions were simulated with 21 Fourier basis for training, and another 100 functions were simulated in the same way for testing. All the simulated functions have the following basis representation
$X_k^{(g)}(t)=\sum_{j=1}^{21}\zeta^{(g)}_{kj}F_j(t),$
where $F_j(t)$ is the $j$-th Fourier basis function. The score vectors of two groups follow the following normal distribution,
$\{\zeta^{(g)}_{kj}\colon{j=1,\ldots,21}\}\sim\mathcal{N}(0,\Sigma_g),$
where $\Sigma_g$ is a diagonal matrix with diagonal elements $\delta_g$. Two settings of $\{\delta_g\colon g=0,1\}$ were considered here, namely, 
\begin{itemize}
\item{Setting 1}: $\delta_0=(1,1,1,0,0,1,1,\ldots,1,1)$, $\delta_1=(1,0,0,1,1,1,1,\ldots,1,1),$
\item{Setting 2}: $\delta_0=(1,1,0,1,0,1,0,1,\ldots,1)$, $\delta_1=(1,0,1,0,1,0,1,1,\ldots,1)$.
\end{itemize}
We applied the VPC to classify the 100 functions in the testing group with finite dimensions $(d=1,\ldots,9)$ and without dimension reduction $(d=\infty)$, and repeated this procedure 200 times. Figure~\ref{d4} displays the average classification rate of the two groups in different settings and the corresponding selected discriminative feature functions. 

\begin{figure}[!h]
\center
\includegraphics[scale=0.45]{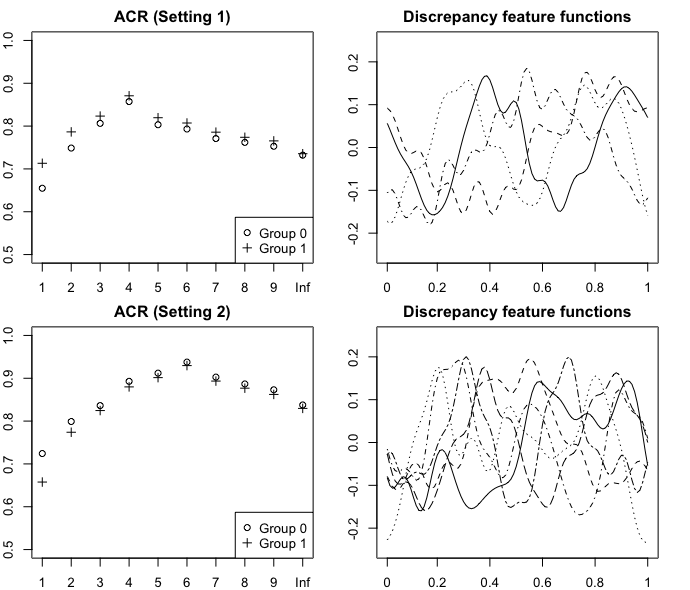}
\caption{Average classification rates under different dimensions, and the corresponding discriminative feature functions. The optimal performance is achieved as $d=4$ (setting 1) and $d=6$ (setting 2) respectively. {``Inf" means that no dimension reduction is implemented before the classification.}}
\label{d4}
\end{figure}

The two groups are differentiated in 4 dimensions in the first setting, and are differentiated in 6 dimensions in the second setting. Figure~\ref{d4} shows that, in these two settings, the classification rate first increases and then decreases as $d$ increases. The optimal performance is achieved as $d=4$ and $d=6$ respectively. Another interesting point is that the discriminative feature functions are similar to those Fourier bases which differentiate the two groups, which provides insights on the discrepancy.

\section{Analysis of Rat Local Field Potentials}
\label{s5}
\subsection{Description of the LFP data and pre-processing}
The VPC method was used to study the impact of a shock (simulated stroke by occlusion) on the 
functional architecture of a rat brain using the local field potential data. Here, LFPs were recorded from multi-tetrodes, implanted on the rat cortex, continuously over a period 6 hours (see Figure~\ref{display}). In this paper, we analyzed a 10-minute window (5 minutes before the occlusion and 5 minutes after the occlusion). 

\begin{figure}[!h]
\center
\includegraphics[scale=0.27]{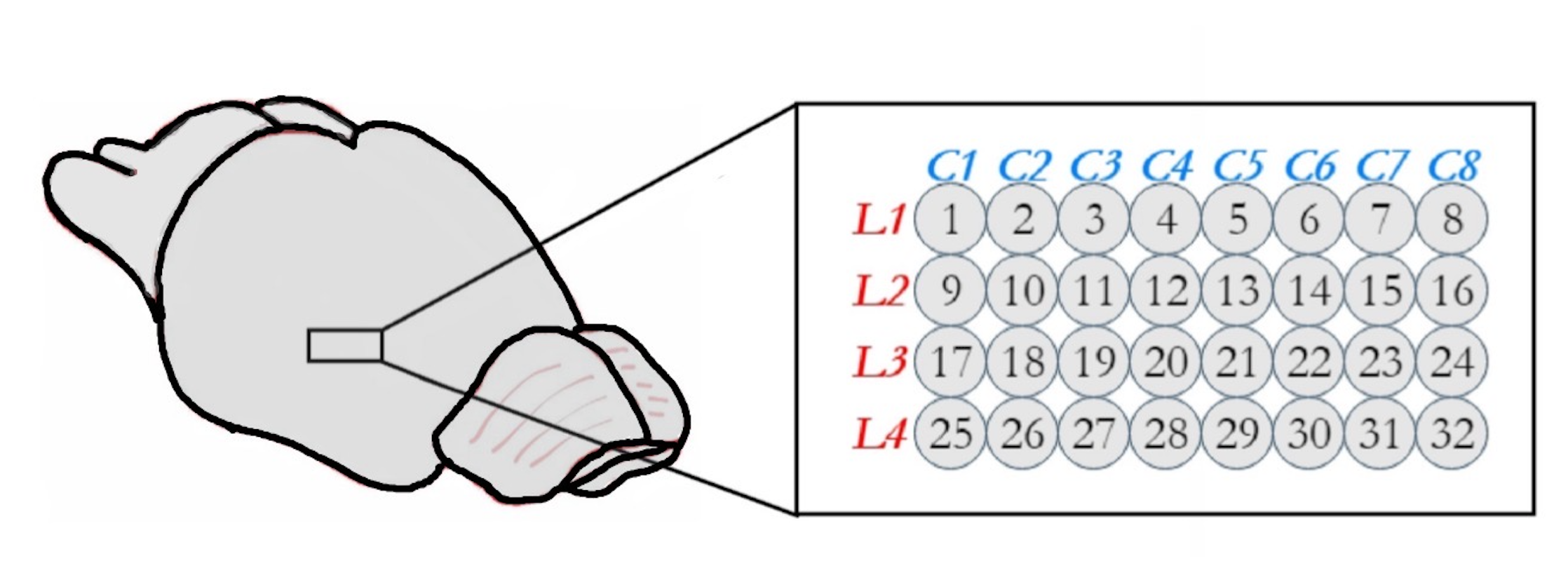}
\caption{Placement of 32 micro-tetrodes, {which was implanted on a patch near the flow field downstream of the clamped brain artery. More details can be found in \cite{r41}.}}
\label{display}
\end{figure}

One important feature of these electrical brain signals is that the means of epochs of different states (pre-stroke and post-stroke onset) are always zero and the interest is mostly focused on the fluctuations or oscillations around the mean. Therefore it is impossible to distinguish different states of brain from the mean difference of these records. We applied the VPC to discriminate between pre-stroke onset ($\Pi_0$) and post-stroke onset ($\Pi_1$) local field potentials.

From these micro-tetrodes, LFPs were recorded at the rate of 1000 observations per second (thus there are $T = 1000$ time points per 1-second epoch). A total of $r = 600$ epochs were recorded. Midway in this period (at epoch $r = 300$), stroke was mechanically induced on the rat by clamping the medial cerebral artery. We extracted the component in different frequency bands from each epoch and used the filtered trajectories in 6 different frequency bands ((1).\ 1-4 Hertz, (2).\ 5-9 Hertz, (3).\ 10-14 Hertz, (4).\ 15-19 Hertz, (5).\ 20-24 Hertz, (6).\ 25-30 Hertz) for classification. The filtered trajectories are obtained by functional smoothing with Fourier basis. For example, to extract the $k$-Hertz component, {one can} smooth the trajectories with the following two Fourier basis functions
\begin{equation*}
\left\{
\begin{array}{cccl}
\sqrt{{2}}\cos(2\pi {k}t), & t\in[0,1],\\
\sqrt{{2}}\sin(2\pi {k}t), & t\in[0,1].
\end{array} \right. 
\end{equation*}
The goals of the analysis are:  (1) to identify features that best differentiate the pre-stroke from the post-stroke onset signals using training data (group labels are known) and (2) to classify a future brain signal because early identification of post-stroke onset signals leads to better treatment outcomes. 

For this particular dataset, the discrepancy was expressed primarily at lag $0$ and thus we did not consider other lags. Data visualization reveals several outliers with extreme amplitude in the post-stroke onset group, which can be easily identified as  post-stroke onset epochs, so we removed them from the samples. All the rest functions (epochs) are scaled to unit norm to achieve balanced classification.

\subsection{Change-point detection and classification with change-points}
Since brain signals can be non-stationary within each group, we considered the structural breaks in the classification. Our approach is to approximate the non-stationary functional sequence as a concatenation of piecewise-stationary functional sequence. Suppose there are $L_0$ and $L_1$ structural breaks in covariance operator dividing the entire sequence of the two groups $(g=0,1)$ into $L_0+1$ and $L_1+1$ stationary segments respectively. Hence, the covariance operator ${C}^{(0)}_{g,k}(\cdot)=E\{\langle X^{(g)}_{k},\cdot\rangle X_{k}^{(g)}\}$ is modeled as follows,
${C}^{(0)}_{g,k}(\cdot)=\sum\limits_{\ell=1}^{L_g+1}\mathbb{I}_{gk}^{(\ell)}\widetilde{C}^{(0)}_{g,\ell}(\cdot),$
where $\widetilde{C}_{g,\ell}^{(0)}(\cdot)$ is the covariance operator of the $\ell$-th quasi-stationary segment of group $g$, and $\mathbb{I}_{gk}^{(\ell)}=1$ if $k$ lies in the $\ell$-th sengment of group $g$ and $\mathbb{I}_{gk}^{(\ell)}=0$ otherwise. 

Here we applied the break point detection method developed in \cite{r24}. Binary segmentation was applied to determine all pronounced break points. The break points of the 32 micro-tetrodes were searched simultaneously, say, a set of common break points were found for the pre-stroke/post-stroke onset epoch sequences of the 32 tetrodes. Let $X^{(g)}_{ki}(\Omega,t)$ be the $\Omega$-frequency component  of the $k$-th epoch obtained from the $i$-th tetrode, and define $$Z^{(g)}_k(\Omega,t,s)=\sum^{32}\limits_{i=1}X^{(g)}_{ki}(\Omega,t)X^{(g)}_{ki}(\Omega,s).$$ 
The break detection method was applied to the sequence $Z^{(g)}_k(\Omega,t,s)$. Under significance level $\alpha=0.05$, several break points were detected in different frequency bands and are displayed in Figure~\ref{cp}. The period of epochs in the same local stationary sequence segmented by the estimated structural break points are shown in the same color.  
\begin{figure}[!h]
\center
\includegraphics[scale=0.35]{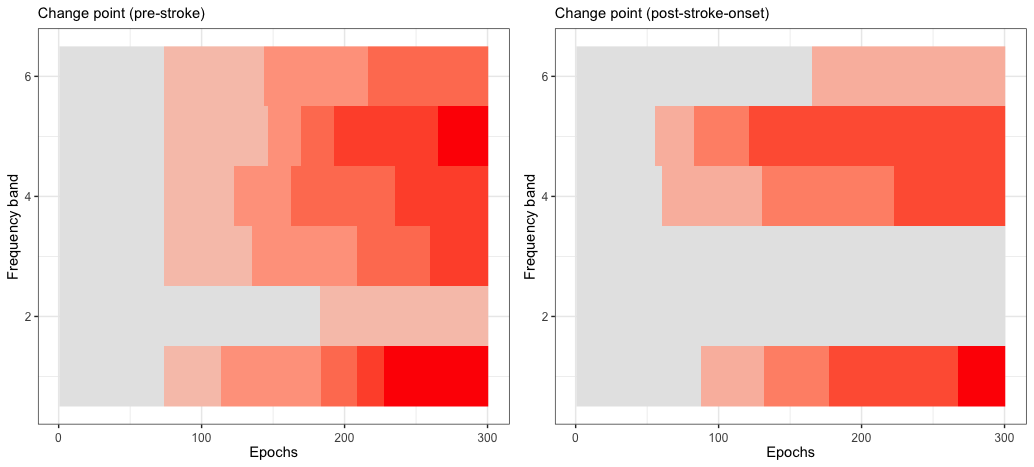}
\caption{Local stationary blocks of pre-stroke and post-stroke onset epochs for different frequency bands. }
\label{cp}
\end{figure}
For each local stationary block, we concatenated the 32 tetrodes' epochs by stacking the observations and obtained the estimated auto-covariance operator function of the concatenated epochs, which served as the classification features. The covariance operators of the concatenated epochs were estimated for all segments. Given a new concatenated function $Y(t)$ with unknown group label, first find the covariance operator closest to $\mathcal{Y}(\cdot)=Y\langle Y,\cdot\rangle$ with resect to the Hilbert-Schmidt norm within each group, denoted as $\widehat{C}^{(0)}_{0,opt}(\cdot)$ and $\widehat{C}^{(0)}_{1,opt}(\cdot)$. Then $Y(t)$ is assigned to group $g$ if the functional operator $\widehat{C}^{(0)}_{g,opt}(\cdot)$ is closer to $\mathcal{Y}(\cdot)$ with respect to the Hilbert-Schmidt norm. See Algorithm~\ref{alg5} for the summary of the classification procedure.

\begin{algorithm}
\caption{Classification algorithm with change-points}
\label{alg5}
\textbf{Step 1}. For each group $(g=0,1)$, apply break point detection method (e.g., \cite{r24}) to find the structural break points in the covariance operator, and estimate the covariance operator for each local stationary subsequence.

\textbf{Step 2}. For each group $(g=0,1)$, find the estimated covariance operator $\widehat{C}^{(0)}_{g,opt}(\cdot)$ that is closest to $\mathcal{Y}(\cdot)$ with respect to the Hilbert-Schmidt norm. Fix $d$, obtain the eigenfunctions $(\widehat{\nu}_{opt,j}\colon j=1,\ldots,d)$ of the functional operator $(\widehat{C}^{(0)}_{0,opt}-\widehat{C}^{(0)}_{1,opt})^2(\cdot)$ associated with its first $d$ largest eigenvalues.

\textbf{Step 3}. Compute the scores $\widehat{S}^{(g)}_{opt,ij}=\langle \widehat{C}^{(0)}_{g,opt}(\widehat{\nu}_{opt,i}),\widehat{\nu}_{opt,j}\rangle$ for $i,j=1,\ldots,d$.

\textbf{Step 4}. Compute $$\widehat{D}_{g,opt}=\sum^{d}_{i,j=1}(\widehat{S}_{opt,ij}^{(g)}-\langle Y,\widehat{\nu}_{opt,i}\rangle\langle Y,\widehat{\nu}_{opt,j}\rangle)^2$$ If $\widehat{D}_{0,opt}-\widehat{D}_{1,opt}<0$, classify $Y$ to $\Pi_0$, otherwise, classify $Y$ to $\Pi_1$.
\end{algorithm}

\subsection{Discriminative feature functions}
To check the overall discriminative features between the two different brain states, the entire pre-stroke and post-stroke onset epochs were employed to calculate the most discriminative feature functions for different frequency bands (1-4 Hertz, 5-9 Hertz, 10-14 Hertz, 15-19 Hertz, 20-24 Hertz, 25-30 Hertz).
\begin{figure}[!h]
\center
\includegraphics[scale=0.31]{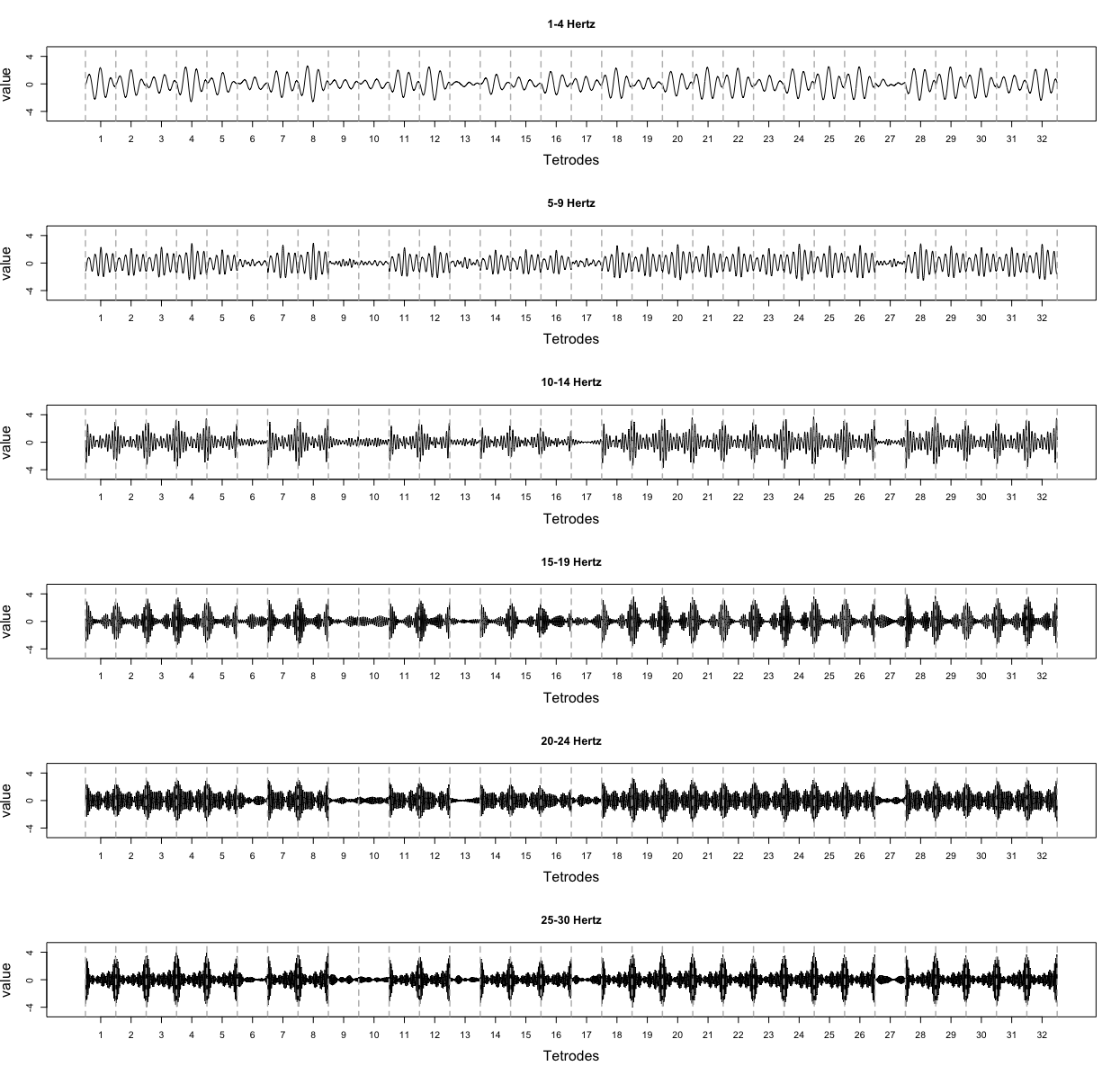}
\caption{The first discriminative feature function of different frequency bands.}
\label{f8}
\end{figure}
Figure~\ref{f8} displays the first discriminative feature functions that best illuminate the discrepancy between the covariance operators of the concatenated epochs for different frequency bands. Each function has 32 equal-length blocks corresponding to 32 tetrodes.

The discriminative functions are informative. First, they reveal the tetrodes which have low power of discrimination. The blocks with low amplitude indicate low power of discrimination of the corresponding tetrodes. For example, it shows that the discriminative feature functions over blocks 9, 10, 13, 17 and 27 are comparatively flat, which indicates that the discrepancy between the two brain states across these tetrodes are comparatively trivial.  Second, they reveal the period in epoch that best differentiates the two states. An interesting finding is that for lower frequency bands (1--9 Hertz), the discrepancy is mainly present in the middle part of epochs, and for higher frequency bands (10--30 Hertz), the discrepancy is mainly present {near the start and ending points of epochs}.

\subsection{Classification performance}
The discrepancy of covariance structure between the two states of brain signals may vary across different tetrodes. Thus we applied the VPC method to each tetrode separately across different frequency bands. A Monte-Carlo procedure was applied. Specifically, at each step, 250 curves were randomly selected as the training set and the rest curves were treated as the testing set.  The Monte-Carlo procedure was repeated 200 times, and the average classification rates (ACR) are displayed in Figure~\ref{twacr}.
\begin{figure}[!h]
\center
\includegraphics[scale=0.35]{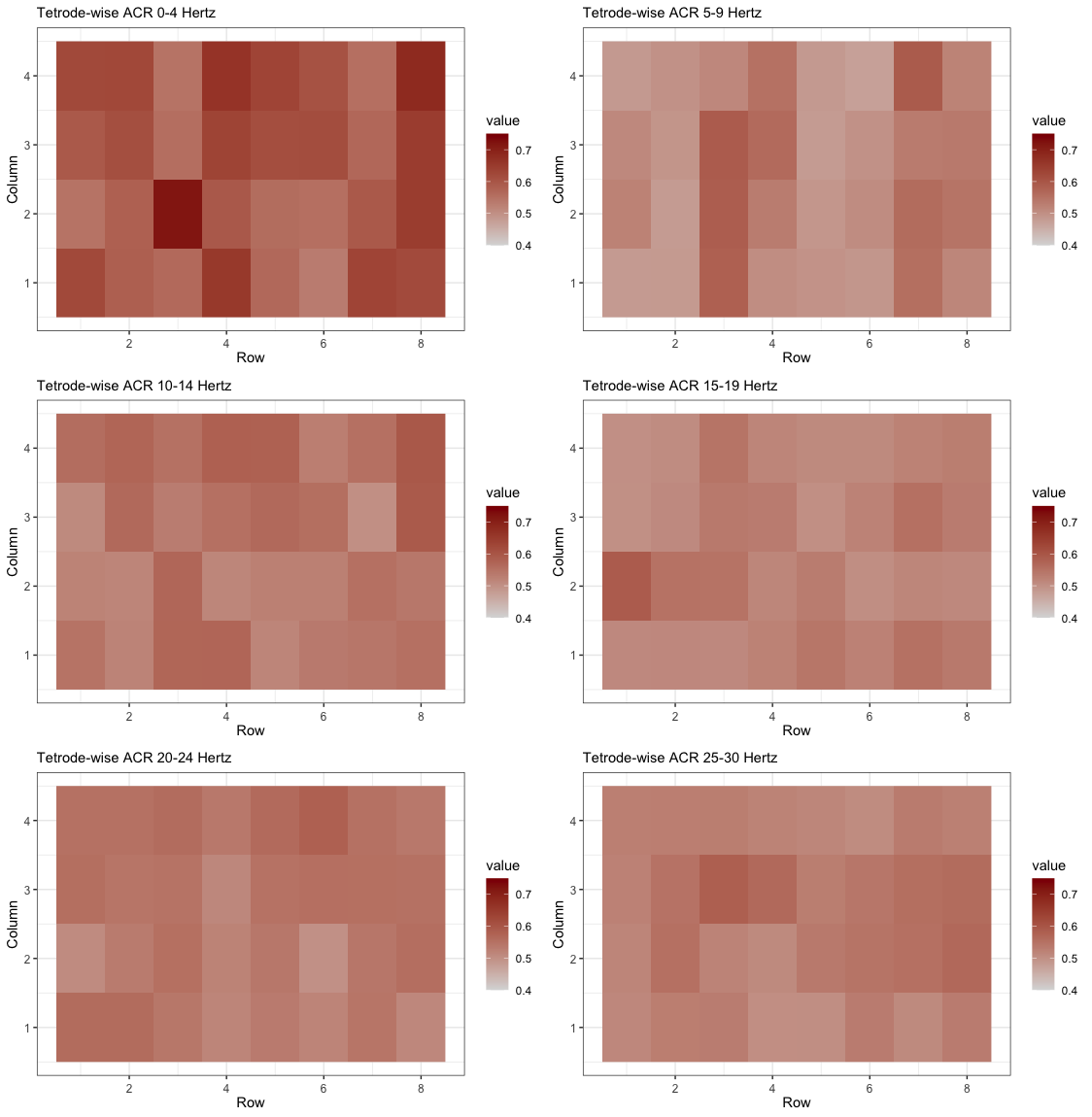}
\caption{The average classification rate based on each individual tetrode. The first row corresponds to tetrodes 1--8, the second row corresponds to tetrodes 9--16, and etc.}
\label{twacr}
\end{figure}

Each subfigure of Figure~\ref{twacr} presents the tetrode-wise average classification rates of one frequency band. An interesting phenomenon is that the $\delta$-band (0--4 Hertz) have the highest power of discrimination. The final classification was conducted on the concatenated epoch trajectories. The same Monte-Carlo procedure was repeated, and the average classification rate of different frequency bands are shown in Table~\ref{t5} together with the standard error of the empirical classification rates. From the table, it is clear that the the classification performance of VPC is {nearly perfect} over 1--9 Hertz, and get worse as the frequency increases. 

\begin{table}[!h]
	\centering
	\caption{Average classification rates and the standard errors (in the parenthesis) across different frequency bands}
	\begin{tabular}{p{0.3in}p{0.5in}p{0.5in}p{0.5in}p{0.5in}p{0.5in}p{0.5in}}
	\hline
	\hline
	\multicolumn{1}{c}{Frequency}&\multicolumn{2}{c}{1--4 Hertz}& \multicolumn{2}{c}{5--9 Hertz}& \multicolumn{2}{c}{10--14 Hertz}\\
	\hline
         State  & $\Pi_0$ & $\Pi_1$ & $\Pi_0$ & $\Pi_1$ & $\Pi_0$ & $\Pi_1$ \\
         \hline
         ACR &0.990 &0.999 &0.980 &0.975 & 0.874&0.970\\
         SE &(0.015)&(0.002) &(0.017) &(0.020) & (0.047)&(0.026)\\
	\hline
	\multicolumn{1}{c}{Frequency}&\multicolumn{2}{c}{15--19 Hertz}& \multicolumn{2}{c}{20--24 Hertz}& \multicolumn{2}{c}{25--30 Hertz}\\
	\hline
         State  & $\Pi_0$ & $\Pi_1$ & $\Pi_0$ & $\Pi_1$ & $\Pi_0$ & $\Pi_1$\\
         \hline 
         ACR&0.898 &0.961 & 0.906&0.919 & 0.841&0.870\\
         SE&(0.043) &(0.027) &(0.041) &(0.043) & (0.051)&(0.045)\\
	 \hline 
	 \hline 
	\end{tabular}
		\label{t5}
\end{table}

\section{Conclusion}
\label{s6}
We developed a general functional approach (VPC) to discriminate and classify different groups of functions with similar means. The VPC method takes advantage of the divergence of the (auto-)covariance operators. The comparison is constrained in the subspace spanned by the discriminative feature functions that account for most of the divergence. The VPC method has a built-in dimension reduction step which extract the major information of discrepancy. Therefore it overcomes the disadvantage of group-wise functional principal components, which only explain variation for each individual group, but do not explain the difference between groups. In addition to improving the classification effectiveness, the discriminative feature functions also reveal the features that differentiate  groups and thus provide insight for classification. We have shown that, the proposed classifier can be perfect under some regularity conditions.

In the application, our methodology provides a general picture of the discrepancy between pre-stroke and post-stroke onset epochs. Specifically, the method reveals the tetrodes, frequencies, and features of epochs that effectively illuminate the discrepancy between different brain states. The VPC has potential medical impact because, if implemented in an online setting, it can help quickly detect the onset of stroke-related abnormal brain electrical rhythms and hence minimize the debilitating effects of stroke. It is  also added that the method has a wide range of applicability including speech signals which is demonstrated in the supplementary material.


\end{document}